\documentclass[lettersize,journal]{IEEEtran}
\usepackage{amsmath,amsfonts}
\usepackage{algorithmic}
\usepackage{algorithm}
\usepackage{array}
\usepackage[caption=false,font=normalsize,labelfont=sf,textfont=sf]{subfig}
\usepackage{textcomp}
\usepackage{stfloats}
\usepackage{url}
\usepackage{verbatim}
\usepackage{graphicx}
\usepackage{cite}
\usepackage{bm} 
\usepackage{booktabs} 
\usepackage{tabularx} 
\usepackage{multirow} 
\newcolumntype{Y}{>{\centering\arraybackslash}X}
\usepackage{xcolor} 
\usepackage[pagebackref=false, colorlinks=true]{hyperref}
\definecolor{IEEEblue}{RGB}{0, 0, 150} 
\hypersetup{
    linkcolor=IEEEblue,   
    citecolor=IEEEblue,  
    urlcolor=IEEEblue,    
}

\begin{document}
\bstctlcite{IEEEexample:BSTcontrol}
\title{A Continuous High Precision Localization Method for AGVs Based on Interferometric FBG Array Sensing and Hybrid Deep Learning}

\author{Laiming~Wu,
        Xin~Gui,
        Hui~Cao,~\IEEEmembership{Member,~IEEE},
        Yunchuan~Zhang,~\IEEEmembership{Member,~IEEE},\\
        Haiyang~Wang,
        Jiaqi~Wang,
        and~Zhengying~Li,~\IEEEmembership{Member,~IEEE}
\thanks{This work was supported by the National Natural Science Foundation of China (Grant No. 62471347), Hubei Provincial Department of Science and Technology (Grant No. 2024BAB022) and the Open Fund (2024KF-01) of Hubei Longzhong Laboratory. (\textit{Corresponding authors: Hui Cao, Yunchuan Zhang.})}%
\thanks{Laiming Wu, Hui Cao, and Yunchuan Zhang are with Hubei Longzhong Laboratory, Wuhan University of Technology Xiangyang Demonstration Zone, Xiangyang 441000, China, and also with the School of Information Engineering, Wuhan University of Technology, Wuhan 430070, China (e-mail: \{315379, iehuicao, yunchuan.zhang\}@whut.edu.cn).}%
\thanks{Xin Gui and Jiaqi Wang are with Hubei Longzhong Laboratory, Wuhan University of Technology Xiangyang Demonstration Zone, Xiangyang 441000, China, and also with the National Engineering Research Center of Fiber Optic Sensing Technology and Networks, Wuhan University of Technology, Wuhan 430070, China (e-mail: \{guixin, wjq\}@whut.edu.cn).}%
\thanks{Haiyang Wang is with Jiangsu Zhiren Jinghang New Materials Research Institute Company Ltd., Nanjing, Jiangsu 210001, China, and also with the School of Information Engineering, Wuhan University of Technology, Wuhan 430070, China (e-mail: wanghaiyang@zrjh.net).}%
\thanks{Zhengying Li is with the National Engineering Research Center of Fiber Optic Sensing Technology and Networks, the State Key Laboratory of Advanced Technology for Materials Synthesis and Processing, and the School of Information Engineering, Wuhan University of Technology, Wuhan 430070, China (e-mail: zhyli@whut.edu.cn).}%
 }



\maketitle

\begin{abstract}
High precision continuous localization is a fundamental challenge for automated guided vehicles (AGVs) operating in electromagnetically sensitive industrial environments. To mitigate the continuity limitations of conventional discrete localization methods and the accuracy limitations of existing continuous localization methods, this paper proposes a continuous AGV localization method that integrates interferometric fiber Bragg grating (FBG) array sensing with a hybrid deep learning framework. We first design a shallowly buried interferometric FBG array sensing track to enable the stable acquisition of high-quality signals. The received sensing signals are subsequently parameterized and preprocessed in a task-oriented manner, in which a multichannel joint representation is constructed using a sliding time window. A hybrid deep learning framework is developed to estimate AGV position from multichannel FBG response signals by jointly exploiting inter-channel spatial correlation and temporal continuity in the measurement sequence. Experimental results show that the proposed method achieves a mean absolute error (MAE) of less than 2 cm under complex industrial conditions, proving its effectiveness and robustness in practical AGV localization scenarios. 
\end{abstract}


\section{Introduction}
\subsection{Context and Motivation}
\IEEEPARstart{A}{utomated} guided vehicles (AGVs) play a vital role in intelligent manufacturing systems and are widely deployed in unmanned factories, intelligent warehousing, and rail logistics. The level of achieved localization accuracy contributes to the operational safety, task scheduling efficiency, and the quality of service (QoS) in the automated system \cite{ref1, ref2}. However, conventional localization methods based on radio frequency (RF) or visual perception often become ineffective in industrial scenarios with high electromagnetic interference and stringent information security constraints. These practical challenges call for a localization strategy capable of providing accurate, continuous, and environmentally robust AGV positioning in the underlying industrial deployments.

Existing AGV localization methods mainly include laser simultaneous localization and mapping (SLAM), vision or QR code based approaches, and magnetic nail or strip based schemes, each with inherent limitations. Among these, laser SLAM constructs environmental maps using LiDAR and achieves autonomous localization through either scan matching \cite{ref3} or probabilistic filtering \cite{ref4}, which has been widely applied to indoor localization scenarios \cite{ref5}. Laser SLAM can achieve high localization accuracy under suitable conditions. Nevertheless, in the target industrial scenarios discussed above, long-term online localization should reduce continuous reliance on environmental scanning and map maintenance. LiDAR based SLAM may also suffer performance degradation under dust contamination, dynamic occlusion, reflective surfaces, sparse or changing geometric features, and workspace structural changes, while its hardware integration and computational cost further increase deployment complexity \cite{ref6, ref7}.

In contrast, typical discrete localization schemes including vision or QR code based approaches and magnetic nail or strip based schemes demonstrate relatively high positioning accuracy with simpler implementations. However, these methods can only provide reliable position updates at discrete reference points, making it difficult to achieve continuous localization along the entire trajectory. Position estimation between adjacent markers usually relies on odometry or interpolation, thus accumulating excessive long-term error \cite{ref8}. Furthermore, these methods often involve high maintenance and adaptation costs in industrial settings \cite{ref9, ref10}. 

\subsection{Related Work}
Distributed optical fiber sensing (DOFS) techniques emerge as a promising strategy for addressing localization challenges \cite{ref11}. Unlike laser SLAM and discrete localization methods, optical fiber sensors offer inherent immunity to electromagnetic interference, high sensitivity, capability for long distance continuous deployment, and ease of multiplexing \cite{ref12}. Existing DOFS-based approaches can be broadly classified into two categories: fully distributed sensing methods based on optical fiber scattering, such as Rayleigh, Raman, and Brillouin scattering \cite{ref13}, \cite{ref14}, \cite{ref15}; and quasi-distributed sensing methods based on fiber Bragg gratings (FBGs) \cite{ref16}. Although sensing methods based on inherent scattering enable truly distributed continuous sensing, their signal to noise ratio (SNR) is typically limited by the inherently low scattering coefficient of the fiber, thereby constraining the sensitivity and spatial resolution of the sensing system \cite{ref17, ref18}. By contrast, sensing based on FBG arrays obtains higher SNR and directly depicts the relationship between sensing responses and physical locations, making it more suitable for high precision distributed detection and localization \cite{ref19, ref20}. 

In strain monitoring based on DOFS, wavelength demodulation \cite{ref21} and phase sensitive optical time domain reflectometry ($ \varphi $-OTDR) \cite{ref22} are two major demodulation techniques. The former one demodulates strain by tracking shifts in the central wavelength of the grating reflection spectrum, but it becomes infeasible when dealing with weak dynamic strain signals. By contrast, $ \varphi $-OTDR reconstructs strain information by extracting phase variations from Rayleigh backscattered signals, showing higher sensitivity to weak dynamic disturbances and thereby providing a better alternative for light load AGV localization \cite{ref23, ref24}. The major concern of $ \varphi $-OTDR lies in its restricted spatial resolution, which is determined by the optical pulse width and typically remains at the meter level, e.g., 3 or 5 m, making it inadequate for continuous high precision AGV localization in unmanned factory environments \cite{ref25, ref26}. Therefore, lifting the limitations in spatial resolution of conventional methods while preserving the high sensitivity of optical fiber sensing remains a fundamental challenge for AGV localization.

In a related line of work, \cite{ref27} developed a mobile road monitoring platform integrated with FBG sensors for vehicle speed measurement. Ultra-weak fiber Bragg grating (UWFBG) sensing technology was adopted to construct a buried optical cable detection system in highway subgrades, in which the sensing fiber was deployed nearly perpendicular to the traffic direction, enabling identification and measurement of traffic flow, speed, and axle spacing in real time \cite{ref28}. As shown recently in \cite{ref29}, a UWFBG array was used to establish a long distance continuous vibration monitoring system for highways, achieving vehicle trajectory detection in real time under complex traffic conditions with a spatial resolution of 15 m. Overall, most of these studies focus on traffic sensing and vehicle state assessment in road traffic scenarios. A related grating array fiber optic AGV positioning study further indicates the feasibility of optical fiber based AGV positioning. However, when an FBG array is deployed along the AGV travel direction, existing feature point or discrete correction based methods still have difficulty providing continuous high precision localization along the entire path \cite{ref30}.

Recent advances in machine learning for optical fiber sensing have shown considerable potential for intelligent signal processing and sensing interpretation \cite{ref31, ref32}. In FBG based sensing, reference \cite{ref33} employed an artificial neural network (ANN) for simultaneous strain and temperature measurement, while authors in \cite{ref34} proposed the ADPNet deep network for strain and temperature decoupling. For distributed fiber optic sensing, a variational autoencoder based representation learning method was introduced for automatic interpretation of strain distributions, enabling crack identification, localization, quantification, and visualization \cite{ref35}. Meanwhile, prior work in \cite{ref36} developed a modified YOLO based framework with transfer learning for real time monitoring of spatially distributed cracks. Deep learning was also adopted in \cite{ref37} and \cite{ref38} for Rayleigh scattering response prediction and phase demodulation of distributed acoustic sensing signals. Moreover, previous studies in \cite{ref39} and \cite{ref40} extended data-driven methods to class incremental signal recognition and transportation infrastructure monitoring in distributed fiber sensing systems. Overall, these studies validate the possibility of data-driven methods in signal demodulation, sensing interpretation, and structural monitoring, but they do not consider continuous localization of a moving AGV from multichannel FBG response signals in enclosed factories. Although this problem remains unexplored in optical fiber sensing, deep learning-based localization frameworks have demonstrated strong potential for vehicle navigation with heterogeneous measurements \cite{ref41, ref42}. Therefore, applying data-driven deep learning methods to model the spatiotemporal heterogeneity of multichannel strain responses induced by a light load AGV can be a promising strategy.

\subsection{Main Contributions}
In this paper, we introduce a novel continuous high precision AGV localization method that combines interferometric FBG array sensing with a model-based deep learning strategy. The main contributions of this work are summarized as follows: 

\begin{itemize}
    \item We develop an interferometric FBG array sensing track buried at a shallow depth to collect stable multichannel response signals induced by AGV motion for building a reliable sensing infrastructure for continuous localization.
    \item We propose a unified AGV localization framework by combining a parametric signal model, a spatiotemporal feature tensor representation, and a hybrid deep learning architecture, thereby enabling effective characterization of waveform morphology, spatial correlation, and temporal continuity in multichannel FBG response signals.
    \item We present experimental results across synthetic tasks and real-world examples. The empirical results reveal that the proposed localization method can achieve a mean absolute error (MAE) of less than 2 cm under complex industrial conditions, confirming strong robustness and promising engineering potential.
\end{itemize}

The remainder of this paper is organized as follows. Section \ref{sec:FBG Array-Based AGV Localization System} presents the AGV localization track based on FBG array sensing, the signal acquisition and demodulation system, and the time domain model of the FBG sensing signals. Section \ref{sec:Deep Learning-Based AGV Localization Method} details the signal processing workflow and the proposed hybrid deep learning framework. Section \ref{sec:Experimental Setup and Performance Evaluation} presents the experimental setup, localization performance evaluation, and ablation and robustness analysis. Finally, Section \ref{sec:Conclusion and Future Work} concludes the paper and discusses future research directions. For better readability, Table \ref{tab:notations} provides a summary of key notations used in this paper.

\begin{table}[!t]
\caption{Summary of Key Notations}
\label{tab:notations}
\centering
\scriptsize
\setlength{\tabcolsep}{2.5pt}
\renewcommand{\arraystretch}{1.12}
\begingroup
\begin{tabularx}{\columnwidth}{@{}>{\centering\arraybackslash}p{0.12\columnwidth}X@{}}
\toprule
\multicolumn{1}{c}{\textbf{Symbol}} & \multicolumn{1}{c}{\textbf{Definition}} \\
\midrule
$F_i(t)$ & Time domain response signal of the $i$-th FBG sensing zone. \\
$\omega$ & Equivalent structural angular frequency used for waveform characterization. \\
$\sigma_{\omega}$, $T_0$ & Effective temporal width and center time of the structural response. \\
$M$ & Number of sensing zones or signal channels. \\
$T$ & Length of the causal sliding time window in samples. \\
$n_k$ & Current sampling index of the $k$-th prediction instant. \\
$\mathbf{x}(n)$ & Multichannel FBG response vector at sampling index $n$. \\
$\tilde{\mathbf{x}}(n)$ & Filtered multichannel FBG response vector. \\
$\mathbf{X}_k$ & Causal temporal sample constructed from current and historical signals. \\
$y_k$, $\hat{y}_k$ & Reference and predicted positions along the predefined sensing route. \\
$\bar{\mathbf{X}}_k$ & Response matrix after amplitude scaling and clipping. \\
$\mathbf{V}_k$, $\mathbf{A}_k$ & First order and second order temporal difference feature matrices. \\
$E_k$ & Average signal energy of the $k$-th temporal sample. \\
$g_k$ & Dynamic weighting coefficient for temporal difference features. \\
$\mathbf{Z}_k$ & Multifeature input formed by concatenating response and difference features. \\
$\hat{\mathbf{Z}}_k$ & Standardized multifeature input for network training and inference. \\
$\mathbf{F}_k$ & Convolutional feature sequence extracted from the input sample. \\
$\mathbf{H}_k$ & Temporal feature sequence generated by the LSTM module. \\
$\boldsymbol{h}_{\mathrm{agg}}$ & Aggregated feature vector used for final position regression. \\
\bottomrule
\end{tabularx}
\endgroup
\end{table}

\section{FBG Array-Based AGV Localization System}
\label{sec:FBG Array-Based AGV Localization System}
\subsection{Localization Track Design}
\begin{figure}[!t] 
\centering
\includegraphics[width=0.92\columnwidth, height=0.24\linewidth, keepaspectratio=false]{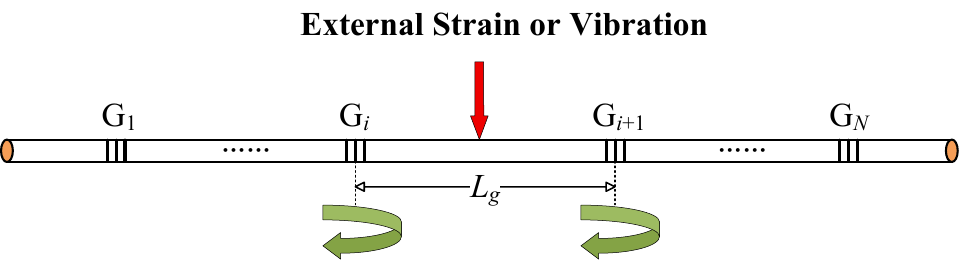}
\caption{Sensing principle of the interferometric FBG array.}
\label{fig_1}
\end{figure}

The sensing principle of the interferometric FBG array is illustrated in Fig. \ref{fig_1}. The reflected light from two adjacent gratings, $G_i$ and $G_{i+1}$, interferes at the demodulation unit, and the corresponding optical path difference (OPD) can be expressed as
\begin{equation}
\mathrm{OPD}_i = 2n_{\text{eff}}L_g,
\label{eq:equation_1}
\end{equation}
where $n_{\text{eff}}$ denotes the effective refractive index of the fiber core, and $L_g$ represents the grating spacing between two adjacent gratings.

According to the optical interference principle, the intensity of the interference signal generated by the $i$-th adjacent grating pair can be expressed as
\begin{equation}
I_{\text{int},i}(t) = I_i + I_{i+1} + 2\sqrt{I_i I_{i+1}} \cos(2kn_{\text{eff}}L_g + \varphi_0 + \Delta\varphi).
\label{eq:equation_2}
\end{equation}
Here, $I_i$ and $I_{i+1}$ denote the reflected light intensities of gratings $\text{G}_i$ and $\text{G}_{i+1}$, respectively. The wavenumber $k$ is given by $k = 2\pi / \lambda$, where $\lambda$ is the optical wavelength. $\varphi_0$ and $\Delta\varphi$ represent the initial phase and the phase perturbation induced by external strain or vibration, respectively.

During AGV operation, strain and vibration induced in the ground are transmitted to the optical fiber, causing slight variations in the grating spacing and the effective refractive index. As a result, the phase term of the interference signal varies dynamically. By accurately demodulating the phase variation $\Delta\varphi$, external disturbances can be detected in real time. In industrial applications, the FBG array is usually encapsulated to improve mechanical stability while preserving sensing sensitivity and environmental adaptability.

\begin{figure}[!t] 
\centering
\includegraphics[width=1\columnwidth, height=0.36\linewidth, keepaspectratio=false]{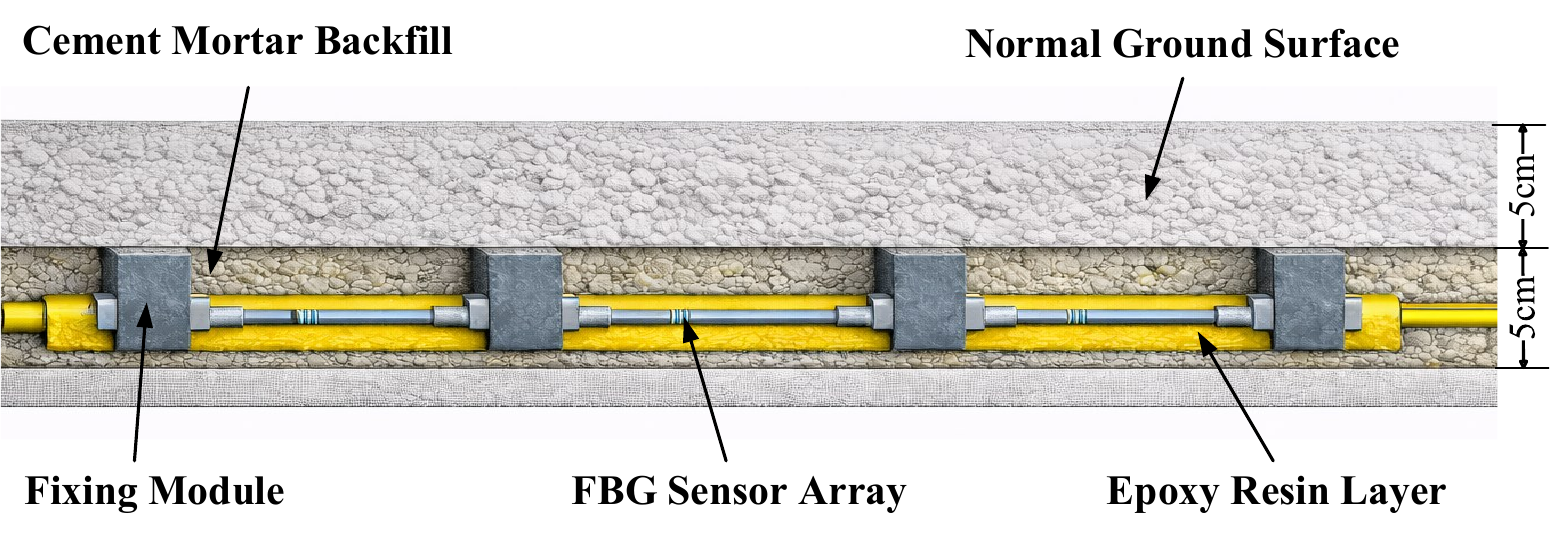}
\caption{Cross-sectional view of the shallowly buried interferometric FBG array sensing track.}
\label{fig_2}
\end{figure}

We design a shallowly buried interferometric FBG array sensing track structure in order to satisfy the AGV localization requirements in unmanned factories under complex operating conditions, as shown in Fig. \ref{fig_2}.

The structure mainly consists of an interferometric FBG array sensing cable, an epoxy resin layer, fixing modules, and cement mortar. The combination of periodic mechanical fixation, cement mortar backfill, and the epoxy resin layer provides stable mechanical coupling between the sensing cable and the surrounding structure. This coupling establishes a continuous and stable strain-transfer path for the response induced by AGV motion. In particular, the epoxy resin layer suppresses interfacial slip and microscale mechanical decoupling between the sensing cable and the backfill medium. This improves strain transfer efficiency and spatial consistency along the cable. Compared with loose burial approaches, the structure enhances the linear response of the system to low frequency loading and quasi static deformation, thus providing more stable signal input for the localization scheme.

The proposed sensing track is intended for predefined industrial AGV routes rather than arbitrary dynamic paths outside the instrumented area. It is not limited to a strictly straight line, since the FBG sensing cable can be installed along planned straight, curved, or segmented routes; for curved routes, the localization coordinate can be defined as the arc-length along the sensing track. Within this setting, the present study focuses on continuous localization along the predefined sensing route, with the longitudinal route coordinate serving as the current localization target. The lateral offset relative to the route centerline provides complementary information for monitoring route deviation and detecting off-track operation.

In practical deployments, the proposed shallowly buried sensing track is mainly intended for fixed or semi-fixed AGV routes rather than frequently changing temporary routes. To improve maintainability, the sensing cable can be installed in modular sections, allowing abnormal or damaged sections to be identified from missing or abnormal sensing responses and replaced locally. For scalability, additional FBG-array segments and reserved shallow grooves can be introduced along planned routes and connected to the demodulation framework within the system capacity. Therefore, the proposed infrastructure trades route flexibility for long-term stable sensing and is more suitable for stable industrial AGV paths.

\begin{figure}[!t] 
\centering
\includegraphics[width=1\columnwidth, height=0.42\linewidth, keepaspectratio=false]{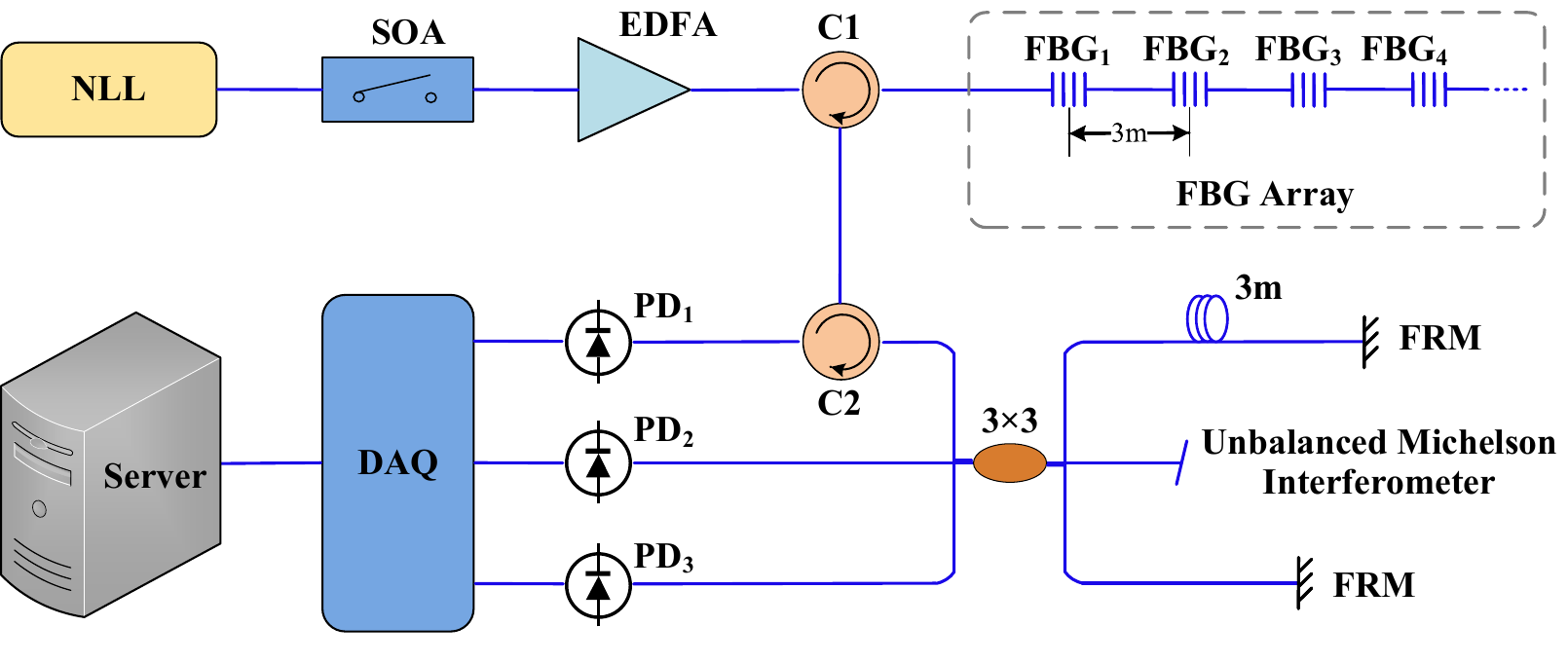}
\caption{Schematic of the signal acquisition and demodulation system based on the interferometric FBG array.}
\label{fig_3}
\end{figure}

\subsection{Signal Acquisition and Demodulation System}
\label{subsec:Signal Acquisition and Demodulation System}
The operating principle of the distributed vibration sensing system based on the FBG array is shown in Fig. \ref{fig_3}.

A narrow-linewidth laser (NLL) with a central wavelength of 1550 nm and a linewidth of 1 kHz serves as the optical source. The continuous wave output from the NLL is modulated by a semiconductor optical amplifier (SOA) to generate an optical pulse train with a repetition rate of 4 kHz and a pulse width of 20 ns. The pulse train is subsequently amplified by an erbium-doped fiber amplifier (EDFA) and routed to the FBG array via Circulator 1 (C1). The array consists of a series of chirped fiber Bragg gratings (CFBGs) with nearly identical parameters, including a central wavelength of 1550 nm, a bandwidth of 5 nm, and a reflectivity of $-46$~dB. Two adjacent CFBGs and the 3 m fiber segment between them together form an independent sensor.

The optical pulses reflected from the FBG array pass through Circulator 2 (C2) and enter a phase demodulation unit based on an unbalanced Michelson interferometer. This unit comprises a $3\times3$ coupler, two Faraday rotation mirrors (FRMs), and three photodetectors (PDs). By demodulating the optical phase variations induced by changes in the optical path length of the fiber segment between adjacent FBGs, the corresponding time domain response signal can be reconstructed. Meanwhile, the disturbed sensing segment can be localized in the time domain according to the optical time domain reflectometry (OTDR) principle. The electrical signals output from the photodetectors are acquired by a data acquisition (DAQ) card and transmitted to a server for subsequent analysis and processing.

\subsection{Modeling of FBG Signals Induced by AGV}
\label{subsec:Modeling of FBG Signals Induced by AGV}
When an AGV travels along the localization track, the signals acquired by the sensing system cannot be directly used for localization modeling and therefore require preprocessing through filtering. Proper selection of the signal filtering band is especially important. The dynamic loading frequencies induced on the track vary with AGV type and operating conditions, depending on factors including wheel size and wheelset configuration, and the traveling speed.

\begin{figure}[!t] 
\centering
\includegraphics[width=1\columnwidth, height=0.36\linewidth, keepaspectratio=false]{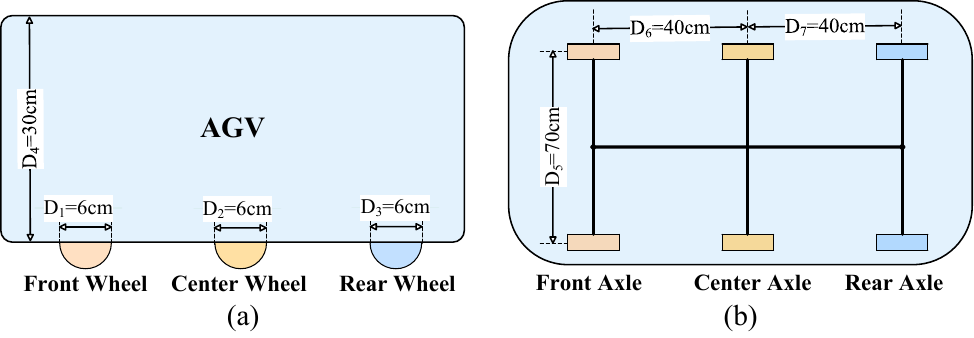}
\caption{Wheel and axle configuration of the AGV.}
\label{fig_4}
\end{figure}

Fig. \ref{fig_4} shows the wheel axle configuration of the AGV used in this study. Specifically, the AGV has three axles and six wheels, and each wheel has a diameter of 6~cm. The axle width is 70~cm. The spacing between adjacent axles is 40~cm. During AGV motion, the wheel and axle loads introduce characteristic dynamic components into the optical fiber sensing signals. We analyze these components to determine the appropriate preprocessing frequency range. Their characteristic frequencies depend on the vehicle speed and the corresponding spatial intervals and can be expressed as
\begin{equation}
f = \frac{v}{L_a},
\label{eq:equation_3}
\end{equation}
with $L_a$ describing the characteristic distance between two adjacent loading points and $v$ denoting the speed of the AGV.

For the equivalent excitation frequency associated with wheel loading, the relevant spatial scale is the wheel--ground contact length $L_w$, which is typically smaller than 1~cm. As a result, the excitation frequency induced by wheel loading remains relatively high even at low AGV speeds. The corresponding frequency exceeds 50~Hz when the traveling speed is 0.5~m/s. Since this frequency range overlaps with high frequency noise in factory environments, it is excluded from further analysis in this work.


\begin{figure}[!t] 
\centering
\includegraphics[width=1\columnwidth, height=0.73\linewidth, keepaspectratio=false]{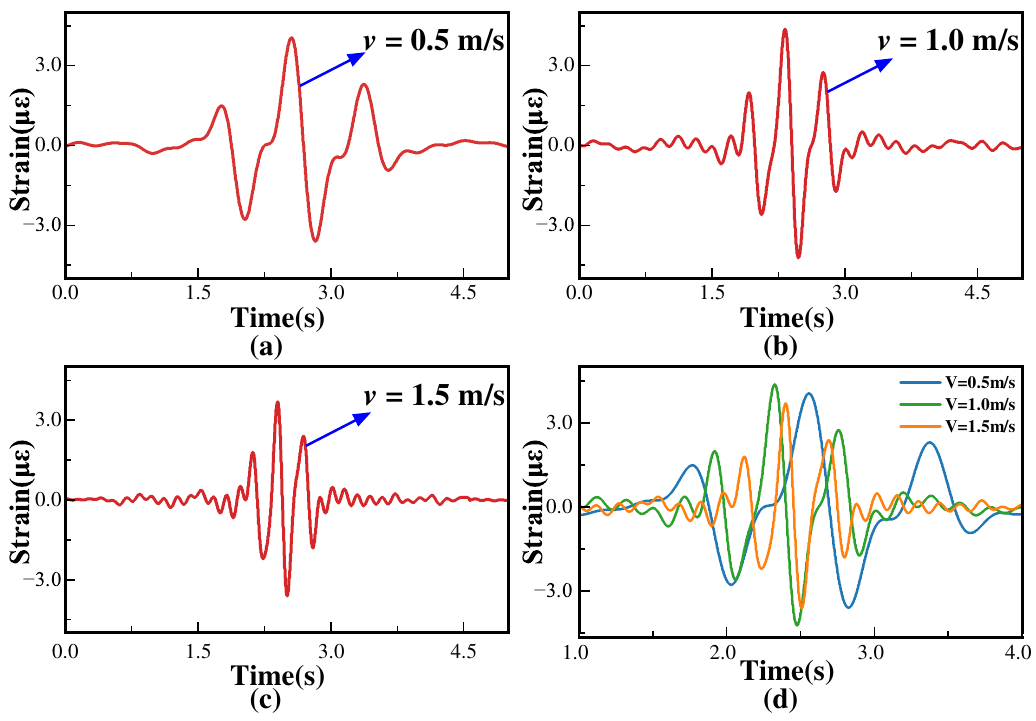}
\caption{Preprocessing results based on the equivalent excitation frequency of axle loading at different AGV speeds: (a) $v = 0.5$ m/s; (b) $v = 1.0$ m/s; (c) $v = 1.5$ m/s; (d) comparison of signals at different speeds.}
\label{fig_5}

\vspace{6pt}

\includegraphics[width=1\columnwidth, height=0.73\linewidth, keepaspectratio=false]{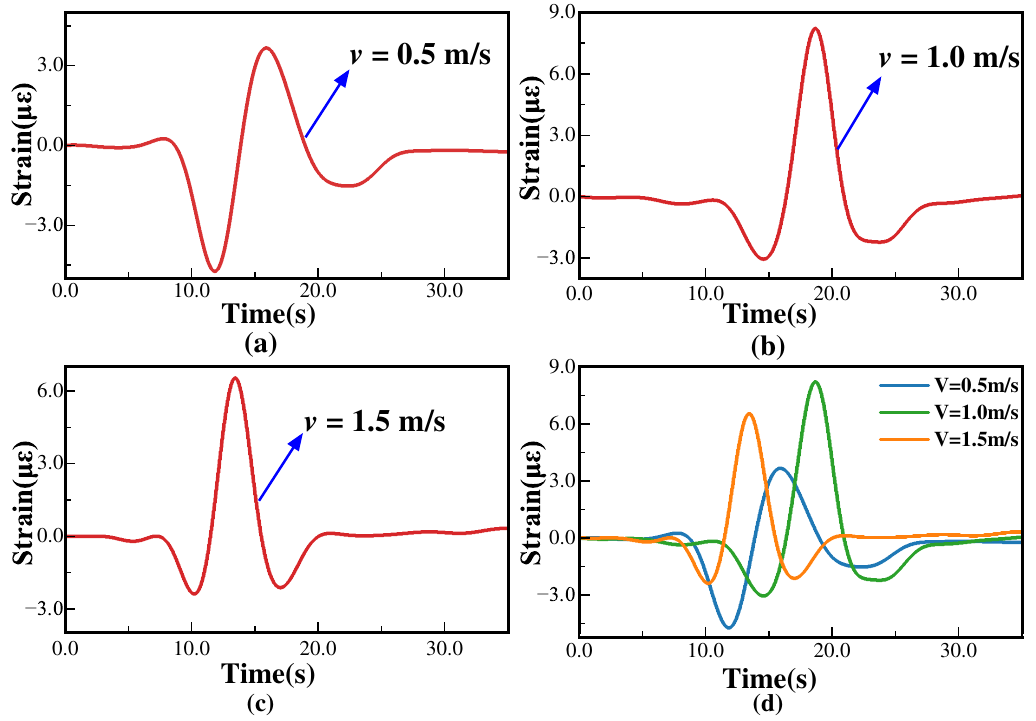}
\caption{Low frequency preprocessing results of the quasi-static response at different AGV speeds: (a) $v = 0.5$ m/s; (b) $v = 1.0$ m/s; (c) $v = 1.5$ m/s; (d) comparison of signals at different speeds.}
\label{fig_6}
\end{figure}

By contrast, for the equivalent excitation frequency with axle loading, when $L_a = 40$~cm, the excitation frequencies are 1.25~Hz, 2.50~Hz, and 3.75~Hz at vehicle speeds of 0.5~m/s, 1.0~m/s, and 1.5~m/s, respectively. The raw signals acquired at different speeds are then filtered within the corresponding frequency bands, and the results are shown in Fig. \ref{fig_5}. It can be observed that the spacing between adjacent axles gives rise to three dominant peaks in the signal waveform. The responses induced by individual axles are superimposed in the time domain. They form a repeated excitation pattern with temporal offsets determined by the axle spacing and vehicle speed. However, different operating speeds require different filtering bands for preprocessing, which increases the complexity of signal processing.

For unified time domain signal modeling, the AGV is treated as an overall moving load. As the vehicle moves through a sensing zone, the load-induced strain varies continuously with the relative position between the AGV and the sensing zone, forming a low-frequency quasi-static response. This response does not require floor-surface irregularities or vehicle-speed fluctuations. The signal components below 1~Hz are therefore extracted using the same low-frequency band for different vehicle speeds, as shown in Fig. \ref{fig_6}. 

Although the signal amplitude and phase vary across sensing zones and vehicle speeds, extensive experimental observations indicate that the time domain waveform presents a stable profile similar to a Gaussian distribution. Based on this feature, we establish a time domain model for AGV induced FBG response signals by statistically fitting a large set of measured data. The model is expressed as
\begin{equation}
F_i(t) = A_i S(t - \Delta t_i) + B_i G(t - \Delta t_i) + n_i(t),
\label{eq:signal_model_4}
\end{equation}
where $i$ denotes the sensing zone index, and $A_i$ and $B_i$ are adjustable amplitude coefficients. Both are assumed to be approximately Gaussian distributed random variables, with distributions given by $A_i \sim \mathcal{N}(\mu_{A_i}, \sigma_{A_i}^2)$, $B_i \sim \mathcal{N}(\mu_{B_i}, \sigma_{B_i}^2)$. In addition, $\Delta t_i$ and $n_i(t)$ represent the temporal shift and the additive noise. The functions $S(t)$ and $G(t)$ describe the intrinsic waveform component and the local excitation envelope, respectively. They are defined as
\begin{equation}
S(t) = \mathrm{sinc}(\frac{\omega}{\pi}(t - T_0)) \exp(-\frac{(t - T_0)^2}{2\sigma_\omega^2})
\label{eq:signal_model_5}
\end{equation}
and
\begin{equation}
G(t) = \exp(-\frac{(t - T_0)^2}{2\sigma^2}).
\end{equation}
Here, $\mathrm{sinc}(\cdot)$ denotes the normalized sinc function, $\omega$ denotes the equivalent structural angular frequency, and $\sigma_\omega$ is the effective temporal window width of the structural response. $T_0$ and $\sigma$ characterize the center time and temporal width of the Gaussian envelope, respectively. The reference standard time domain waveform can be written as
\begin{equation}
F(t) = H_1 S(t) + H_2 G(t).
\label{eq:signal_model_7}
\end{equation}

\begin{figure}[!t] 
\centering
\includegraphics[width=0.86\columnwidth, height=0.51\linewidth, keepaspectratio=false]{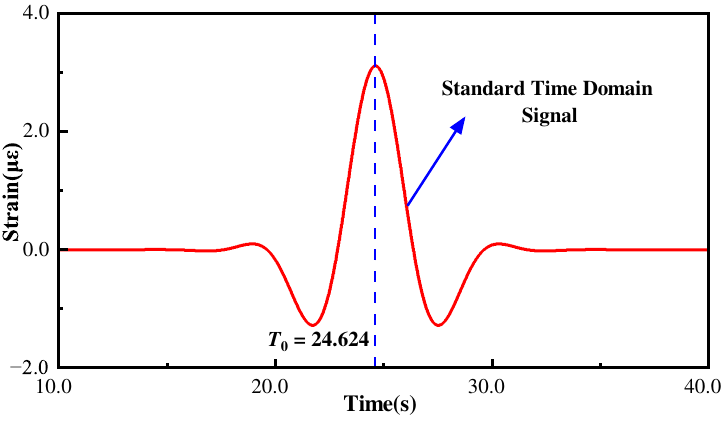}
\caption{The standard time domain waveform.}
\label{fig_7}
\end{figure}

The waveform of $F(t)$ is shown in Fig. \ref{fig_7}, and the corresponding parameter values are listed in Table \ref{tab:estimated_parameters}.

\begin{table}[!t]
\caption{The Estimated Parameters in the Standard \\ Time Domain Signal}
\label{tab:estimated_parameters}
\centering
\footnotesize
\begin{tabularx}{\columnwidth}{YYY}
\toprule
\textbf{Parameter symbol} & \textbf{Value} & \textbf{Unit} \\
\midrule
$H_1$             & 5.1037   & ---   \\ [3pt] 
$H_2$             & -1.9941  & ---   \\ [3pt]
$\omega$          & 1.3329   & rad/s \\ [3pt]
$\sigma_{\omega}$ & 3.1975   & s     \\ [3pt]
$\sigma$          & 2.0037   & s     \\ [3pt]
$T_0$             & 24.624   & s     \\
\bottomrule
\end{tabularx}
\end{table}

\subsection{Problem Formulation}
As above, the temporal response characteristics of AGV motion within a single sensing zone were described using the signal model given in (\ref{eq:signal_model_4})--(\ref{eq:signal_model_7}). However, it remains difficult to achieve continuous AGV localization from an individual sensing zone alone, since the vehicle position is jointly encoded in the synchronized temporal evolution and spatial distribution of the responses across the entire FBG sensor array. Therefore, the localization task in this work is formulated at the multichannel system level.

Let $\mathbf{X}_k \in \mathbb{R}^{M \times T}$ denote the multichannel signal segment acquired from a sliding time window, where $M$ is the number of sensing zones and $T$ is the number of temporal samples in each window. The objective of the localization model is to learn the nonlinear mapping
\begin{equation}
f: \mathbb{R}^{M \times T} \rightarrow \mathbb{R}, \quad y_k = f(\mathbf{X}_k) 
\label{eq:equation_8}
\end{equation}
with $y_k$ representing the reference longitudinal position along the predefined sensing route. For a curved route, $y_k$ can be defined as the arc-length coordinate measured from the route origin. The localization errors reported in this study are calculated with respect to this longitudinal route coordinate.

Accordingly, the predicted position is expressed as
\begin{equation}
\hat{y}_k = f(\mathbf{X}_k),
\label{eq:equation_9}
\end{equation}
where $\hat{y}_k$ is the estimated AGV position of the $k$-th sample.

Therefore, the continuous localization task considered in this work can be regarded as a nonlinear regression problem that maps synchronized multichannel FBG response signals to the corresponding AGV position. Based on this formulation, Section \ref{sec:Deep Learning-Based AGV Localization Method} develops a deep learning framework to learn the above nonlinear mapping from multichannel sensing data.

\section{Deep Learning-Based AGV Localization Method}
\label{sec:Deep Learning-Based AGV Localization Method}
\subsection{Framework Overview}
\begin{figure}[!t] 
\centering
\includegraphics[width=0.92\columnwidth, height=0.3\linewidth, keepaspectratio=false]{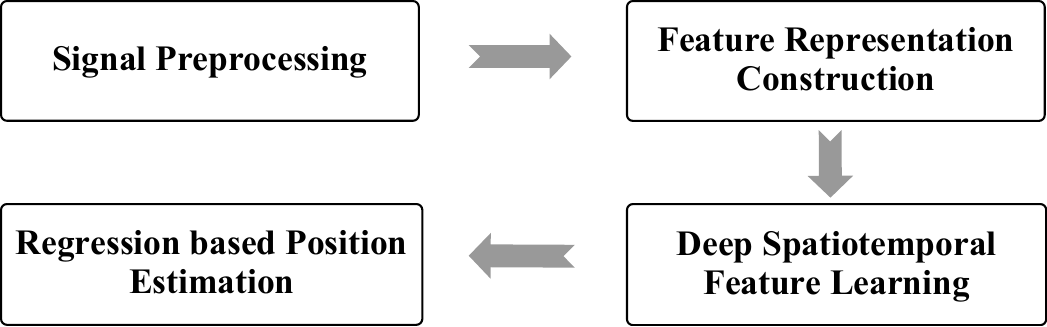}
\caption{Overall workflow of the proposed AGV localization method based on deep learning.}
\label{fig_8}
\end{figure}

We develop a method based on deep learning for continuous AGV localization from synchronized multichannel FBG response signals. As illustrated in Fig. \ref{fig_8}, the proposed method consists of four main stages: signal preprocessing, feature representation construction, deep spatiotemporal feature learning, and regression based position estimation. Together, these stages form a unified pipeline for continuous AGV localization from multichannel FBG response signals.

Within this scheme, raw sensing signals are first transformed into samples suitable for model input through preprocessing and feature representation construction. The resulting feature representations are then processed by a deep network to learn the spatial correlations and temporal dependencies associated with AGV motion. Finally, the learned high-level features are mapped to AGV position estimates through a regression module.

\subsection{Signal Preprocessing and Feature Construction}
\label{subsec:Signal Preprocessing and Feature Construction}
\begin{figure}[!t] 
\centering
\includegraphics[width=0.67\columnwidth, height=1\linewidth, keepaspectratio=false]{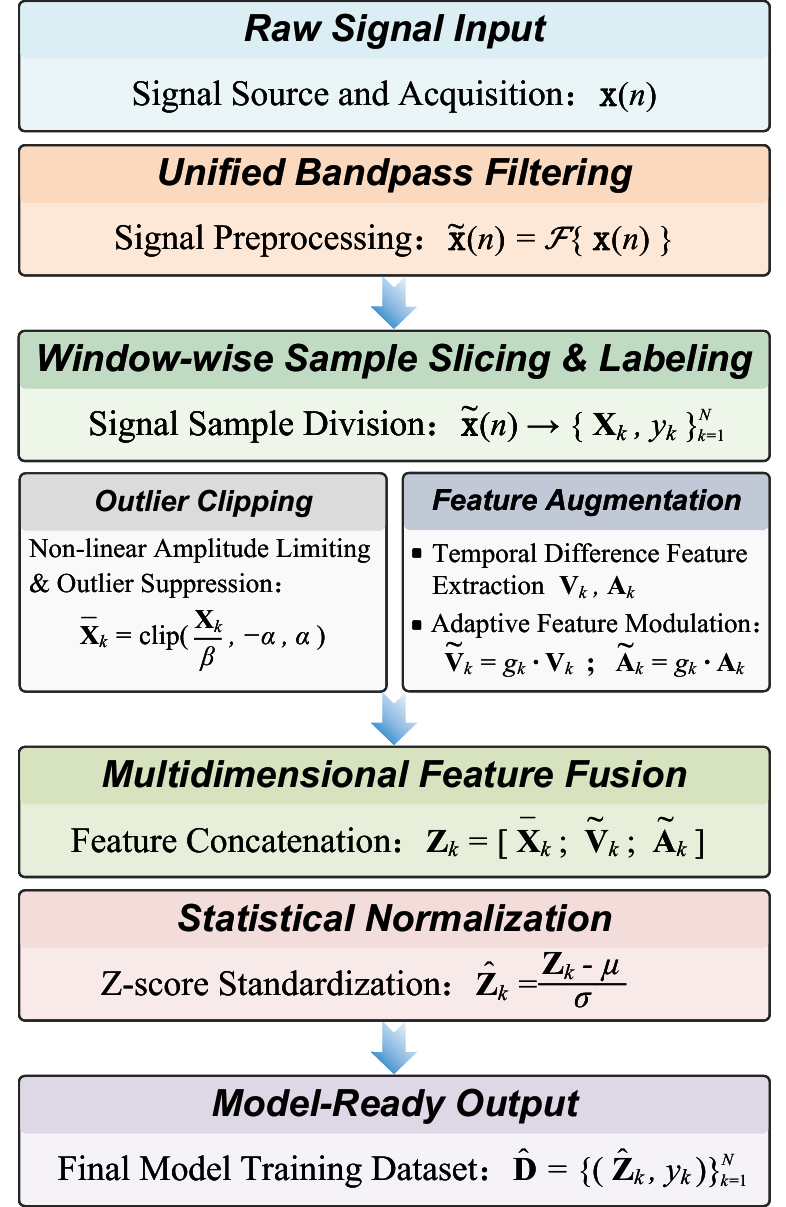}
\caption{Signal preprocessing and feature construction.}
\label{fig_9}
\end{figure}

In practical industrial environments, raw multichannel FBG response signals are inevitably influenced by external disturbances and sensor noise, causing amplitude fluctuations, outliers, and other undesirable variations. Direct use of such signals for network training may lead to unstable optimization and degraded prediction accuracy. Therefore, the raw signals are subjected to preprocessing and feature construction prior to model training. The overall procedure is illustrated in Fig. \ref{fig_9}.

The multichannel response signal of the sensing system at sampling index $n$ can be expressed as
\begin{equation}
\mathbf{x}(n) = [x_1(n), x_2(n), \cdots, x_M(n)]^{\text{T}},
\label{eq:equation_10}
\end{equation}
where $x_m(n)$ denotes the response signal of the $m$-th sensing zone at sampling index $n$. In the present system, each sensing zone corresponds to one FBG sensing channel. The signals are filtered to suppress high frequency noise and remove random interference, yielding
\begin{equation}
\tilde{\mathbf{x}}(n) = \mathcal{F}\{\mathbf{x}(n)\},
\label{eq:equation_11}
\end{equation}
with $\tilde{\mathbf{x}}(n)$ being the filtered multichannel response signals. $\mathcal{F}(\cdot)$ represents the filtering operator. In this work, a bandpass filter is adopted to preserve the effective frequency components of the raw signals.

\begin{figure}[!t] 
\centering
\includegraphics[width=0.83\columnwidth, height=0.70\linewidth, keepaspectratio=false]{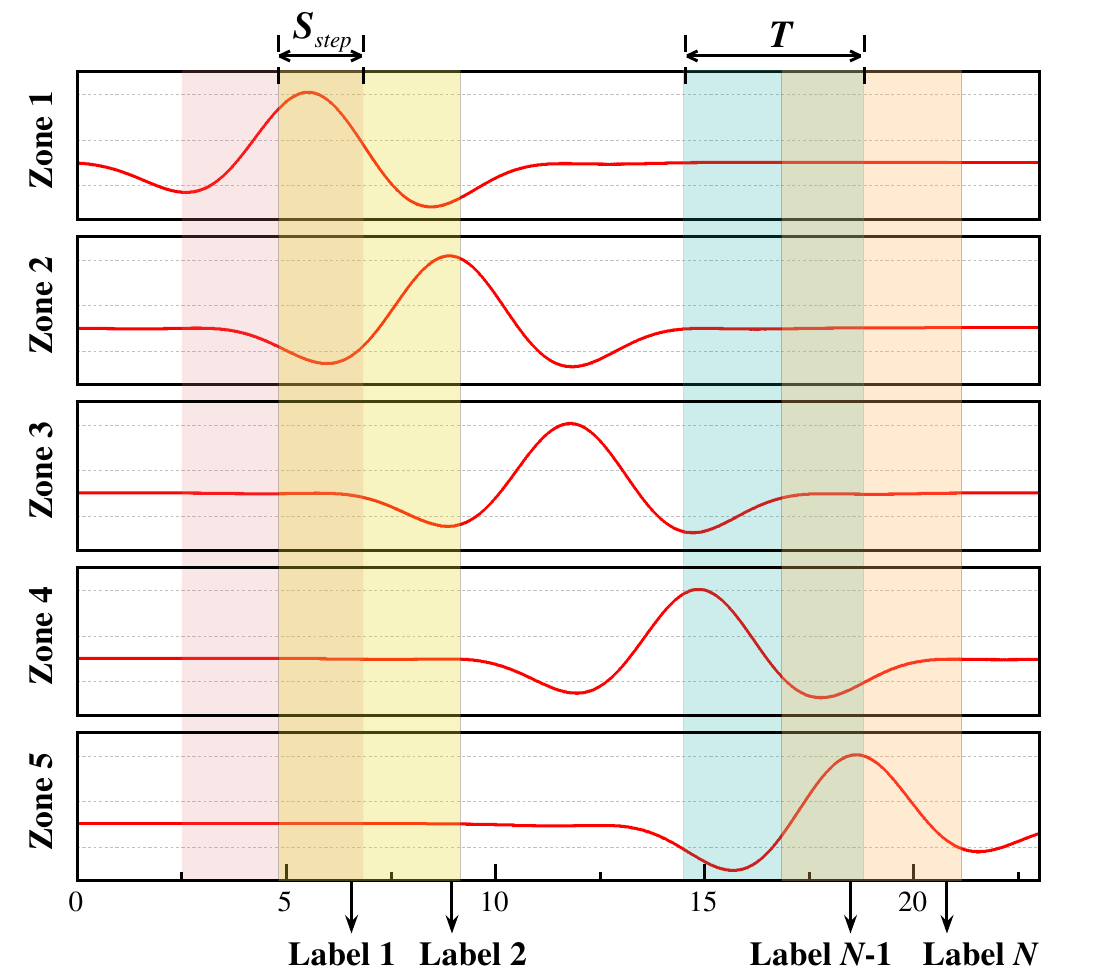}
\caption{Sample segmentation using a right-aligned causal sliding time window.}
\label{fig_10}
\end{figure}

After filtering, a causal sliding time window mechanism is employed to segment the time domain signals of each sensing zone into input samples, as shown in Fig. \ref{fig_10}. At each prediction instant, only the current and historical signal samples are used to construct the model input. Let $n_k$ denote the current sampling index of the $k$-th prediction instant. The corresponding temporal sample is defined as
\begin{equation}
\mathbf{X}_k = [\tilde{\mathbf{x}}(n_k - T + 1), \tilde{\mathbf{x}}(n_k - T + 2), \cdots, \tilde{\mathbf{x}}(n_k)],
\label{eq:equation_12}
\end{equation}
where $T$ denotes the length of the causal sliding time window in samples.
By pairing this sample with the corresponding position label $y_k$, the complete dataset can be constructed as
\begin{equation}
\mathbf{D} = \{(\mathbf{X}_k, y_k)\}_{k=1}^N,
\label{eq:equation_13}
\end{equation}
where $N$ is the total number of samples, and $y_k$ denotes the reference position of the AGV at the current sampling index $n_k$. The reference position labels are generated by a laser SLAM-aided localization system and are time-aligned with the endpoint of each causal input window.

Since the input window is right-aligned with the current prediction instant, the window length specifies the amount of historical context used for feature representation. During online operation, after the initial buffer is available, predictions are updated according to the sliding stride without using future observations. The specific window length, sliding stride, and inference latency are reported in Section \ref{sec:Experimental Setup and Performance Evaluation}.

To suppress possible outliers in the signal, amplitude clipping is applied to the input sample $\mathbf{X}_k$ before temporal difference extraction and feature fusion. The operation can be written as
\begin{equation}
\bar{\mathbf{X}}_k = \mathrm{clip}(\frac{\mathbf{X}_k}{\beta}, -\alpha, \alpha).
\label{eq:equation_14}
\end{equation}
Here, $\beta$ represents the amplitude scaling factor, $\alpha$ is the symmetric clipping threshold, and $\mathrm{clip}(\cdot)$ denotes the amplitude clipping operator.

Considering the direct relationship between the dynamic characteristics of AGV motion and the temporal variation of the response signals, we introduce first and second order temporal difference features to enhance motion related information. By using the same causal sliding time window as in (\ref{eq:equation_12}), the first-order and second-order temporal difference feature matrices are defined as
\begin{equation}
\mathbf{V}_k = [\mathbf{v}(n_k - T + 1), \mathbf{v}(n_k - T + 2), \cdots, \mathbf{v}(n_k)]
\label{eq:equation_15}
\end{equation}
and
\begin{equation}
\mathbf{A}_k = [\mathbf{a}(n_k - T + 1), \mathbf{a}(n_k - T + 2), \cdots, \mathbf{a}(n_k)],
\label{eq:equation_16}
\end{equation}
where $\mathbf{V}_k$ and $\mathbf{A}_k$ denote the first and second order temporal difference feature matrices of the $k$-th sample, respectively. $\mathbf{v}(n)$ and $\mathbf{a}(n)$ represent the first and second order temporal difference features, respectively.

In addition, we introduce a dynamic feature weighting mechanism based on signal energy, because variations in signal energy during AGV motion can reflect the vehicle operating state. The dynamic weighting coefficient is determined based on the energy of sample $\bar{\mathbf{X}}_k$ and is defined as
\begin{equation}
g_k = \mathrm{clip}(\frac{E_k}{\gamma}, g_{\min}, g_{\max}),
\label{eq:equation_17}
\end{equation}
where
\begin{equation}
E_k = \frac{1}{MT} \sum_{i=1}^M \sum_{j=1}^T \left| \bar{\mathbf{X}}_k(i, j) \right|^2
\label{eq:equation_18}
\end{equation}
denotes the average energy of the $k$-th temporal sample, with $M$ representing the number of sensing zones and $T$ representing the number of temporal samples in each causal sliding window. 
Here, $\gamma$ is the energy normalization factor. $g_{\min}$ and $g_{\max}$ represent the lower and upper bounds of the weighting coefficient, respectively. This coefficient is used to adaptively modulate the temporal difference features in (\ref{eq:equation_15}) and (\ref{eq:equation_16}), yielding enhanced features $\tilde{\mathbf{V}}_k = g_k \mathbf{V}_k$ and $\tilde{\mathbf{A}}_k = g_k \mathbf{A}_k$ to strengthen transient vibration information associated with AGV motion. The original clipped response matrix is then concatenated with the enhanced temporal difference features to form a new multifeature input: $\mathbf{Z}_k = [\bar{\mathbf{X}}_k; \tilde{\mathbf{V}}_k; \tilde{\mathbf{A}}_k]$, and $\mathbf{Z}_k \in \mathbb{R}^{C \times T}$, where $C = 3M$.

\begin{figure}[!t] 
\centering
\includegraphics[width=1\columnwidth, height=0.58\linewidth, keepaspectratio=false]{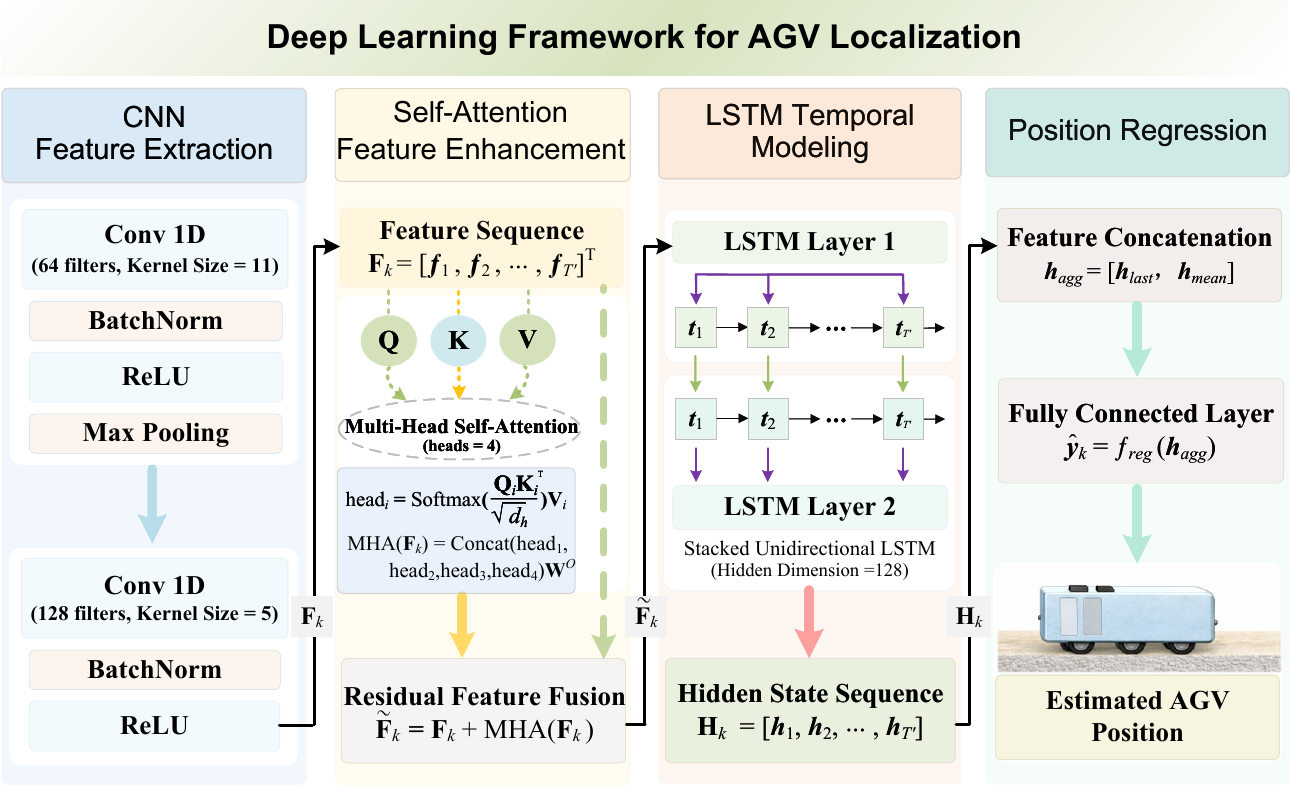}
\caption{Architecture and information flow of the proposed hybrid deep learning network for AGV localization.}
\label{fig_11}
\end{figure}

Finally, the enhanced multifeature input is standardized as
\begin{equation}
\hat{\mathbf{Z}}_k = \frac{\mathbf{Z}_k - \mu}{\sigma},
\label{eq:equation_19}
\end{equation}
where $\mu$ and $\sigma$ denote the mean and standard deviation of the training data, respectively.

After preprocessing and feature construction, the input feature set for model training is obtained as $\hat{\mathbf{D}} = \{(\hat{\mathbf{Z}}_k, y_k)\}_{k=1}^N$. The proposed representation enhances the expressive ability of the input features by preserving the original response information and incorporating dynamic temporal difference features and energy adaptive weighting. This enhanced feature representation provides a more informative and stable input for subsequent deep spatiotemporal feature learning.

\subsection{Hybrid Deep Learning Localization Network}
A hybrid deep learning framework is developed for AGV localization by integrating local pattern extraction with convolutional neural network (CNN), global dependency modeling with self-attention, and temporal sequence learning with long short-term memory (LSTM). As shown in Fig. \ref{fig_11}, the network consists of four main modules: a convolutional feature extraction module, a self-attention feature enhancement module, an LSTM temporal modeling module, and a position regression module. The CNN output feature sequence $\mathbf{F}_k$ is passed to the self-attention module and combined with the attention output through residual fusion to obtain $\widetilde{\mathbf{F}}_k$. The enhanced feature sequence $\widetilde{\mathbf{F}}_k$ is then fed into the LSTM module to generate the hidden-state sequence $\mathbf{H}_k$, which is subsequently aggregated and passed to the position regression module.
By jointly modeling correlations across sensing zones and temporal dependencies in multichannel FBG response signals, the proposed architecture improves localization stability and prediction accuracy under complex operating conditions.

We first employ one-dimensional convolutional operations to extract local temporal patterns from multichannel FBG response signals. The process of feature extraction can be expressed as
\begin{equation}
\mathbf{F}_k^{(1)} = \mathrm{ReLU}(\mathrm{BN}(\mathrm{Conv}_1(\hat{\mathbf{Z}}_k)))
\label{eq:equation_20}
\end{equation}
and
\begin{equation}
\mathbf{F}_k^{(2)} = \mathrm{ReLU}(\mathrm{BN}(\mathrm{Conv}_2(\mathrm{Pool}(\mathbf{F}_k^{(1)})))),
\label{eq:equation_21}
\end{equation}
where $\mathrm{Conv}(\cdot)$ denotes the one-dimensional convolution operation, $\mathrm{BN}(\cdot)$ denotes batch normalization, $\mathrm{ReLU}(\cdot)$ represents the rectified linear unit activation function, and $\mathrm{Pool}(\cdot)$ denotes the max-pooling operation.

Specifically, the first convolutional layer employs 64 convolutional filters with a kernel size of 11. The subsequent max-pooling layer down-samples the temporal dimension, thereby reducing the sequence length and improving feature reliability. The second convolutional layer uses 128 convolutional filters with a kernel size of 5. After convolutional feature extraction, the resulting feature representation is denoted by $\mathbf{F}_k^{(2)} \in \mathbb{R}^{128 \times T'}$, where $T'$ denotes the temporal sequence length after pooling.

To better model long range temporal dependencies, we adopt a multi-head self-attention mechanism. The convolutional feature map is then transposed to arrange the temporal steps along the first dimension and reformulated as a feature sequence: $\mathbf{F}_k = [\bm{f}_1, \bm{f}_2, \cdots, \bm{f}_{T'}]^{\mathrm{T}} \in \mathbb{R}^{T' \times 128}$, where $\bm{f}_t \in \mathbb{R}^{128}$ denotes the feature vector at the $t$-th time step. The multi-head self-attention operation can be defined as
\begin{equation}
\mathrm{head}_i = \mathrm{Softmax}\left(\frac{\mathbf{Q}_i \mathbf{K}_i^{\mathrm{T}}}{\sqrt{d_h}}\right) \mathbf{V}_i, \quad i = 1, 2, \cdots, H
\label{eq:equation_22}
\end{equation}
and
\begin{equation}
\mathrm{MHA}(\mathbf{F}_k) = \mathrm{Concat}(\mathrm{head}_1, \mathrm{head}_2, \cdots, \mathrm{head}_H) \mathbf{W}^O,
\label{eq:equation_23}
\end{equation}
where $\mathbf{Q}_i = \mathbf{F}_k \mathbf{W}_i^Q$, $\mathbf{K}_i = \mathbf{F}_k \mathbf{W}_i^K$, $\mathbf{V}_i = \mathbf{F}_k \mathbf{W}_i^V$, and $\mathbf{W}_i^Q, \mathbf{W}_i^K, \mathbf{W}_i^V \in \mathbb{R}^{128 \times d_h}$ are the learnable projection matrices for the $i$-th attention head. Here, $\mathbf{W}^O$ denotes the output projection matrix, $H$ is the number of attention heads, and $d_h$ is the embedding dimension of each head. Through the multi-head self-attention mechanism, the network can adaptively assign greater importance to temporal segments that are more informative for AGV motion characterization, therefore enhancing global contextual modeling.

In addition, to preserve the original convolutional descriptors while incorporating global contextual information, we adopt a residual fusion strategy, as given by $\tilde{\mathbf{F}}_k = \mathbf{F}_k + \mathrm{MHA}(\mathbf{F}_k)$. The residual formulation enables stable feature propagation in the deep architecture and prevents informative local features from being diluted during feature transformation.

Since the AGV induced FBG response signals show clear temporal continuity and dynamic evolution, the enhanced feature sequence is further processed by an LSTM network for temporal modeling. The hidden state update can be written as
\begin{equation}
\boldsymbol{h}_t = \mathrm{LSTM}(\tilde{\boldsymbol{f}}_t, \boldsymbol{h}_{t-1}),
\label{eq:equation_24}
\end{equation}
where $\tilde{\boldsymbol{f}}_t$ denotes the input feature at time step $t$, and $\boldsymbol{h}_t$ is the corresponding hidden state. In this study, a two-layer stacked LSTM architecture is adopted, with the hidden state dimension set to 128. Considering the need to avoid information leakage from future observations during online inference, we employ a unidirectional LSTM structure. Since each input window is right-aligned with the current sampling index $n_k$, the model only uses the current and historical FBG response signals for position prediction. Therefore, the proposed framework satisfies the causality requirement of real-time AGV localization systems. After LSTM temporal modeling, the resulting hidden state sequence is denoted by $\mathbf{H}_k = [\boldsymbol{h}_1, \boldsymbol{h}_2, \cdots, \boldsymbol{h}_{T'}]$, where $\mathbf{H}_k \in \mathbb{R}^{T' \times 128}$.

\begin{algorithm}[!t]
\caption{Training Process of the Localization Framework}
\label{alg:training_process}
\begin{algorithmic}[1]

\item[\textbf{Input:}] Training set $\mathcal{D}_{\text{train}}$, validation set $\mathcal{D}_{\text{val}}$, epoch number $E$, and batch size $B$
\item[\textbf{Output:}] Optimal network parameters $\Theta^*$

\STATE \textbf{procedure} \textsc{TrainLocalization}($\mathcal{D}_{\text{train}}$, $\mathcal{D}_{\text{val}}$, $E$, $B$)
\STATE Generate model-ready inputs $\hat{\mathbf{Z}}_k$ for all samples.
\STATE Initialize the localization network $f_{\Theta}(\cdot)$, the AdamW optimizer, and the one cycle learning-rate scheduler.
\STATE $\mathrm{MAE}_{\text{best}} \leftarrow +\infty$
\FOR{$\text{epoch} = 1$ \textbf{to} $E$}
    \FOR{each mini-batch $(\hat{\mathbf{Z}}_b, y_b)$ \textbf{in} $\mathcal{D}_{\text{train}}$}
        \STATE $\hat{y}_b \leftarrow f_{\Theta}(\hat{\mathbf{Z}}_b)$
        \STATE Compute $\mathrm{L}_{\text{total}}$ using \eqref{eq:equation_28} and \eqref{eq:equation_29}
        \STATE Backpropagate and clip the gradient norm to 1.0
        \STATE Update $\Theta$ and the learning-rate scheduler
    \ENDFOR
    \STATE $\mathrm{MAE}_{\text{val}} \leftarrow \text{Evaluate}(\mathcal{D}_{\text{val}})$
    \IF{$\mathrm{MAE}_{\text{val}} < \mathrm{MAE}_{\text{best}}$}
        \STATE $\Theta^* \leftarrow \Theta$; $\mathrm{MAE}_{\text{best}} \leftarrow \mathrm{MAE}_{\text{val}}$
    \ENDIF
\ENDFOR
\STATE \textbf{return} $\Theta^*$
\STATE \textbf{end procedure}

\end{algorithmic}
\end{algorithm}

Subsequently, the LSTM output sequence is aggregated to obtain a more compact feature representation with global information. The aggregated feature vector is constructed by feature concatenation as
\begin{equation}
\boldsymbol{h}_{\mathrm{agg}} = [\boldsymbol{h}_{\mathrm{last}}, \boldsymbol{h}_{\mathrm{mean}}] \in \mathbb{R}^{256},
\label{eq:equation_25}
\end{equation}
with $\boldsymbol{h}_{\mathrm{last}} = \boldsymbol{h}_{T'}$ being the hidden state at the final time step of the sequence, and
\begin{equation}
\boldsymbol{h}_{\mathrm{mean}} = \frac{1}{T'} \sum_{i=1}^{T'} \boldsymbol{h}_i
\label{eq:equation_26}
\end{equation}
represents the global feature obtained by average pooling over the entire sequence. Finally, the AGV position is predicted through a multilayer fully connected regression network:
\begin{equation}
\hat{y}_k = f_{\mathrm{reg}}(\boldsymbol{h}_{\mathrm{agg}}),
\label{eq:equation_27}
\end{equation}
with $f_{\mathrm{reg}}(\cdot)$ representing the regression network.

\subsection{Training Strategy and Loss Function}
During training, the network parameters are optimized using the AdamW optimizer. As a variant of Adam, AdamW introduces decoupled weight decay for parameter regularization. This mechanism helps mitigate overfitting and improves model generalization.

For learning rate scheduling, a one cycle policy is adopted to dynamically regulate the optimization process. In the early training stage, the learning rate is gradually increased to encourage broader exploration of the parameter space. As training proceeds, the learning rate is gradually reduced so that the model can approach a better solution through stable parameter updates. This strategy not only stabilizes training but also accelerates convergence, thus reducing the inefficiency commonly associated with fixed learning rate settings. In addition, we apply gradient clipping during training to mitigate exploding gradients in backpropagation, thereby improving optimization stability.

\begin{table}[!t]
\caption{Dataset Composition}
\label{tab:dataset_composition}
\centering
\footnotesize
\begin{tabularx}{\columnwidth}{YYYY}
\toprule
\textbf{Data Type} & \textbf{AGV Speed (m/s)} & \textbf{Load Condition} & \textbf{Number of Samples} \\
\midrule
\begin{tabular}[c]{@{}c@{}}Simulated\\ Signal Data\end{tabular} 
& 1.00 & Unloaded & 29510 \\ 
\midrule 
\multirow{5}{*}{\begin{tabular}[c]{@{}c@{}}Experimental\\ Vibration Data\end{tabular}} 
            & 0.75 & Unloaded & 39510 \\ [3pt]
            & 1.00 & Unloaded & 29510 \\ [3pt]
            & 1.25 & Unloaded & 23510 \\ [3pt]
            & 1.50 & Unloaded & 19510 \\ [3pt]
            & 1.00 & Loaded   & 29510 \\
\bottomrule
\end{tabularx}
\end{table}

\begin{table}[!t]
\caption{Experimental Parameters and Training Configurations}
\label{tab:experimental_parameters_config}
\centering
\footnotesize

\begin{tabularx}{\columnwidth}{>{\hsize=0.8\hsize}Y >{\hsize=1.4\hsize}Y >{\hsize=0.8\hsize}Y}
\toprule
\textbf{Category} & \textbf{Parameter} & \textbf{Value} \\
\midrule

\multirow{5.5}{*}{\begin{tabular}[c]{@{}c@{}}Data\\ Preparation\end{tabular}} 
    & Number of FBG sensing zones & $M = 9$ \\ [3pt]
    & Time window width           & $T = 500$ \\ [3pt]
    & Sample sliding stride       & $S_{\text{step}} = 10$ \\ [3pt]
    & Symmetric clipping threshold & $\alpha = 3.0$ \\ [3pt]
    & Amplitude scaling factor    & $\beta = 0.5$ \\
\midrule

\multirow{6.5}{*}{\begin{tabular}[c]{@{}c@{}}Training\\ Configuration\end{tabular}} 
    & Optimizer               & AdamW \\ [3pt]
    & Learning rate scheduler & OneCycleLR \\ [3pt]
    & Maximum learning rate   & $1\times10^{-3}$ \\ [3pt]
    & Weight decay            & $1\times10^{-2}$ \\ [3pt]
    & Batch size              & 64 samples \\ [3pt]
    & Total epochs            & 200 \\
\midrule

\multirow{2.0}{*}{\begin{tabular}[c]{@{}c@{}}Regularization\\ \& Stabilization\end{tabular}} 
    & Gradient clipping norm  & 1.0 \\ [3pt]
    & Dropout rate            & 0.2 \\
\midrule

\multirow{3.2}{*}{Loss Function} 
    & Huber delta                & $\delta = 1.0$ \\ [3pt]
    & Excess error threshold     & $\tau = 5\text{ cm}$ \\ [3pt]
    & Penalty scaling coefficient & $\lambda = 0.8$ \\
\bottomrule
\end{tabularx}
\end{table}

We design a composite loss function to jointly optimize overall localization accuracy and suppress extreme prediction errors. For a single training sample, the composite loss is defined as
\begin{equation}
\mathrm{L}_{\mathrm{total}} = \mathrm{L}_{\mathrm{Huber}} + \lambda [\max(0, |y - \hat{y}| - \tau)]^3,
\label{eq:equation_28}
\end{equation}
where $\mathrm{L}_{\mathrm{Huber}}$ denotes the basic regression loss, $\tau$ is the excess error threshold, and $\lambda$ is the penalty scaling coefficient for the excess error term. Compared with mean squared error (MSE), the Huber loss is more robust to outlier samples. When the prediction error is small, the Huber loss behaves quadratically, allowing the model to achieve accurate fitting. Once the error exceeds the Huber switching threshold, the loss transitions to linear growth, reducing the influence of abnormal samples on model training. It can be defined as
\begin{equation}
\mathrm{L}_{\mathrm{Huber}}(y, \hat{y}) = 
\begin{cases}
\frac{1}{2}(y - \hat{y})^2, & \text{for } |y - \hat{y}| \le \delta \\[8pt]
\delta |y - \hat{y}| - \frac{1}{2}\delta^2, & \text{otherwise}
\end{cases}
\label{eq:equation_29}
\end{equation}
where $y$ and $\hat{y}$ denote the ground truth and predicted physical positions of the AGV, respectively, and $\delta$ is the switching threshold.

The detailed training process of our proposed localization framework is illustrated in Algorithm \ref{alg:training_process}.

\begin{figure}[!t] 
\centering
\includegraphics[width=1\columnwidth, height=0.45\linewidth, keepaspectratio=false]{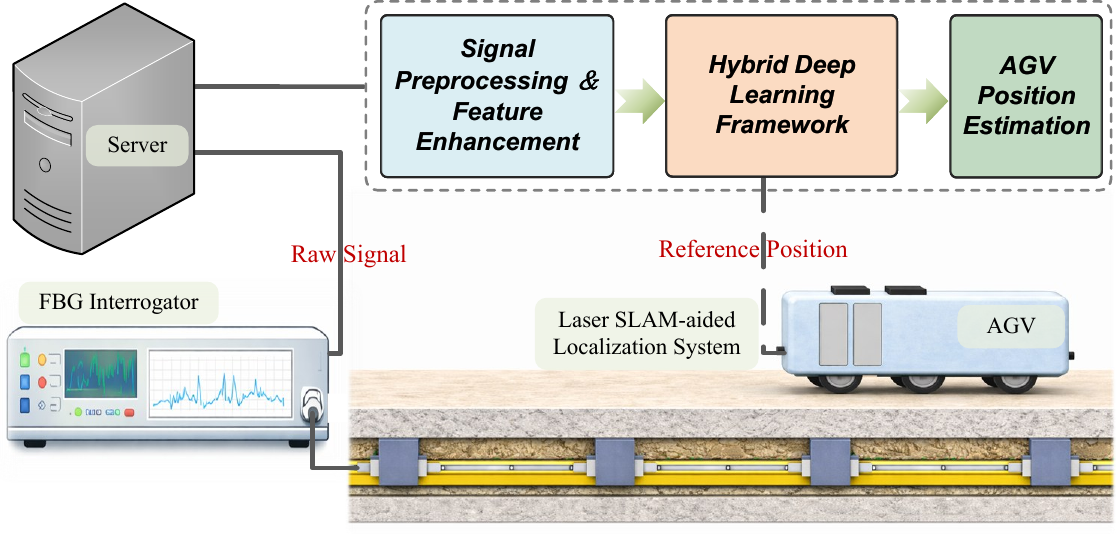}
\caption{Experimental platform.}
\label{fig_12}
\end{figure}

\section{Experimental Setup and Performance Evaluation}
\label{sec:Experimental Setup and Performance Evaluation}
\subsection{Experimental Platform}
To verify the feasibility of the proposed AGV localization method, an experimental platform was established, as shown in Fig. \ref{fig_12}. The platform consists of an FBG sensing track, an FBG array interrogation system, and the proposed hybrid deep learning localization framework. It is used for AGV motion sensing and localization performance evaluation.

FBG array sensors were shallowly embedded within the track and mechanically coupled to the track structure through periodic fixation and an epoxy resin bonding layer. When the AGV travels along the track, the structural strain induced by its motion is transferred to the FBG array sensors. The FBG interrogator performs signal acquisition and demodulation in real time at a sampling rate of 1000~Hz. The demodulated raw signals are then transmitted to a server for subsequent processing. All experiments reported in this study were conducted using the same epoxy-bonded sensing track installation described above.

In addition, the laser SLAM-aided localization system mounted on the AGV is used only as an offline reference trajectory source for dataset construction and performance evaluation. The system uses a Furui H1E0-03B navigation LiDAR and adopts a natural contour localization algorithm. Before dataset construction, the SLAM trajectory is aligned with the FBG sensing track coordinate system and temporally synchronized with the FBG sensing signals to generate reference position labels. The aligned SLAM positions were verified using a laser distance meter, with the positioning deviation within $\pm$5~mm. During online inference, the trained model estimates the AGV position directly from multichannel FBG response signals without using the laser SLAM-aided system.

\subsection{Dataset Construction and Parameter Settings}
We construct a dataset consisting of simulated signals and experimentally acquired vibration signals, as summarized in Table \ref{tab:dataset_composition}. The two types of signals are used at different stages of model development and evaluation.

Before deploying the localization model on the experimental hardware platform, we first generate a dataset of simulated signals according to the FBG strain signal model established in Section \ref{subsec:Modeling of FBG Signals Induced by AGV} [Eq. (\ref{eq:signal_model_4})] for model pretraining. The simulated condition corresponds to an AGV speed of 1.00~m/s and a load condition of 300~kg. Under controlled noise conditions, this dataset is used to assess the feature extraction ability of the proposed hybrid deep learning framework.

To better evaluate the localization performance of the model under practical operating conditions, we collect high fidelity vibration signals from the AGV at different vehicle speeds (0.75--1.50~m/s) and load conditions (unloaded 300~kg and loaded 500~kg). The collected signals are then processed through the uniform bandpass filtering and window-wise sample slicing and labeling procedures introduced in Section \ref{subsec:Signal Preprocessing and Feature Construction} to construct the experimental vibration dataset. Finally, for each operating condition, the resulting samples are randomly divided into training, validation, and test sets with proportions of 80\%, 10\%, and 10\%, respectively.

Moreover, the detailed experimental parameters and training configurations are summarized in Table \ref{tab:experimental_parameters_config}.

\emph{Real-time inference setting:}
With the experimental configuration of $T=500$, a sampling rate of 1000~Hz, and a sample sliding stride of 10 samples, each causal input window contains 0.5~s of historical FBG response signals, while the steady-state prediction update interval is 10~ms. The proposed model contains 473,601 trainable parameters. The inference test was conducted on a platform equipped with an NVIDIA GeForce RTX 4050 Laptop GPU, and the average single-inference latency of the neural network was 0.32~ms. Therefore, the measured inference latency accounts for only about 3.2\% of the update interval, indicating that the proposed framework is computationally feasible for online AGV localization on a GPU-assisted computing platform.


\begin{figure}[!t] 
\centering
\includegraphics[width=1\columnwidth, height=0.6\linewidth, keepaspectratio=false]{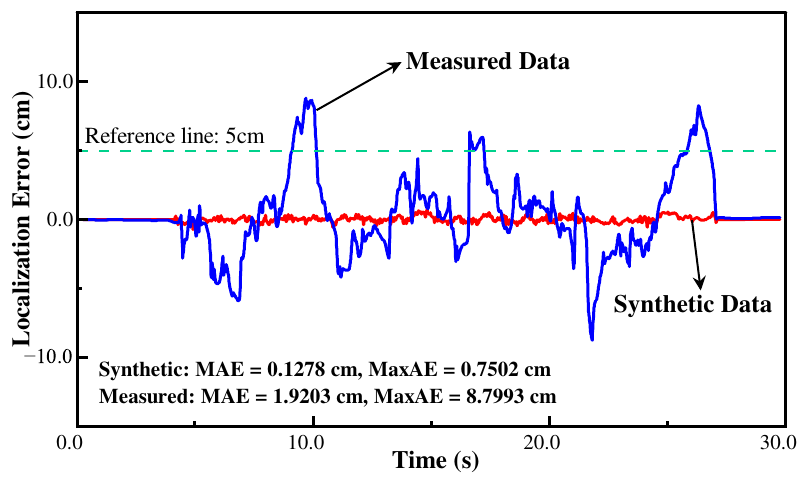}
\caption{Localization errors for synthetic and measured signals.}
\label{fig_13}

\vspace{6pt}

\includegraphics[width=1.02\columnwidth, height=0.86\linewidth, keepaspectratio=false]{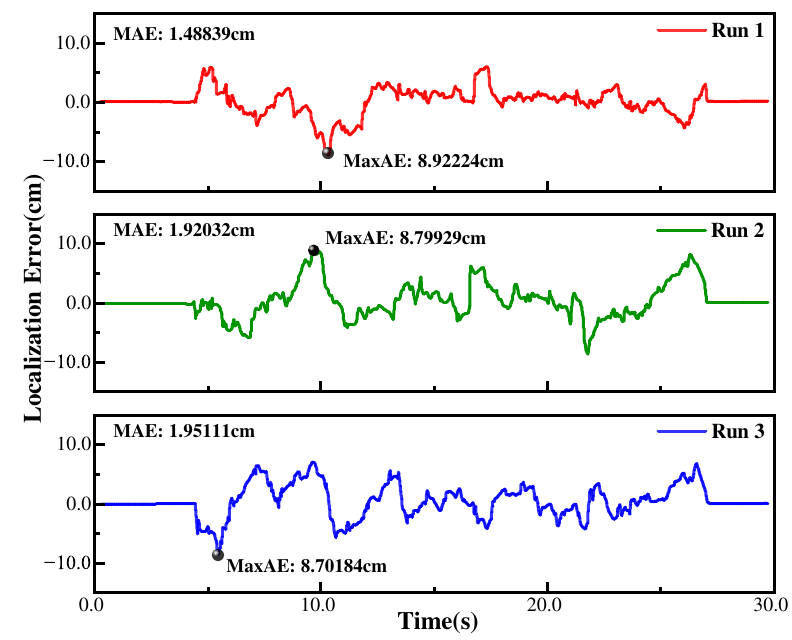}
\caption{Repeated experimental results at the same operating speed.}
\label{fig_14}
\end{figure}

\begin{figure*}[!t]
\centering
\includegraphics[width=1\textwidth]{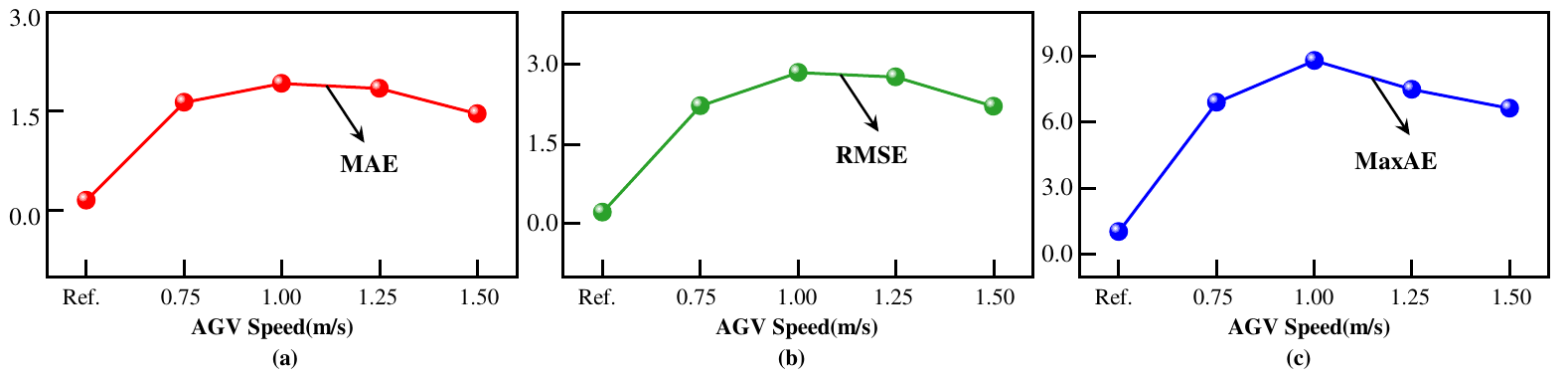} 
\caption{Localization results under different speed conditions.}
\label{fig_15} 
\end{figure*}

\subsection{Localization Performance Evaluation}
Based on the dataset and parameter settings described in the previous subsection, the proposed hybrid deep learning framework is trained and then applied to AGV localization. The localization errors are computed by comparing the predicted positions with the corresponding ground truth positions.

Fig. \ref{fig_13} presents the localization errors for the synthetic signals and the measured signals. Under the synthetic signal condition, the proposed framework achieves an MAE of 0.1278~cm and a maximum absolute error (MaxAE) of 0.7502~cm, indicating that the network can accurately learn the mapping between the signal features and AGV position under ideal conditions. For the measured signals, the MAE and MaxAE are 1.9203~cm and 8.7993~cm, respectively. Although the localization error increases due to environmental disturbances, measurement noise, and structural uncertainties in the practical system, the average localization error remains at the centimeter level. Compared with the published FBG based AGV positioning work in \cite{ref30}, which reported an average driving positioning error of about 2.41~cm, the proposed method achieves competitive accuracy while providing continuous position estimation from multichannel FBG response sequences. This result verifies the effectiveness of the proposed method for continuous AGV localization in practical applications.

\begin{figure}[!t] 
\centering
\includegraphics[width=1\columnwidth, height=0.6\linewidth, keepaspectratio=false]{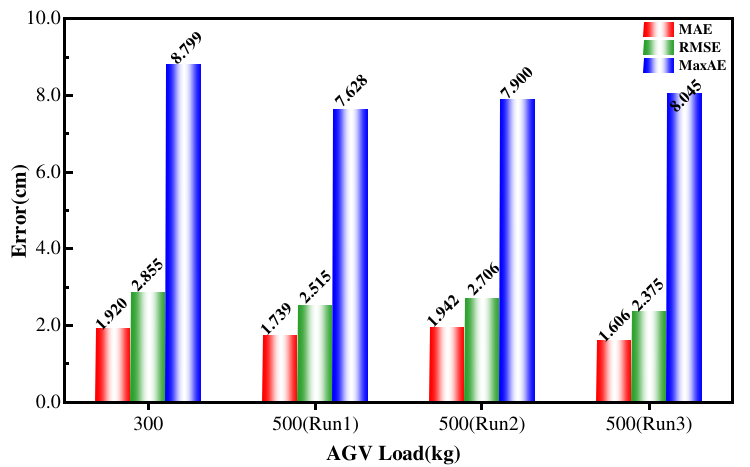}
\caption{Localization results under different vehicle mass conditions.}
\label{fig_16}
\end{figure}

Repeated experiments are conducted under the same speed condition to evaluate the repeatability of the proposed method. As shown in Fig. \ref{fig_14}, the MAE values of the three repeated trials are 1.48839~cm, 1.92032~cm, and 1.95111~cm, respectively, while the corresponding MaxAE values are 8.92224~cm, 8.79929~cm, and 8.70184~cm. The high consistency among these results confirms that the proposed method has good repeatability and stable localization performance under identical operating conditions. This consistency also indicates that the learned spatiotemporal representation is not dominated by occasional fluctuations in individual experimental runs.

The stability of the proposed method to speed variation is evaluated at different AGV speeds. As summarized in Fig. \ref{fig_15}, compared with the reference condition, the localization error increases at practical operating speeds but remains relatively stable. Specifically, for AGV speeds of 0.75~m/s, 1.00~m/s, 1.25~m/s, and 1.50~m/s, the MAE remains below 2~cm, while the MaxAE is consistently below 9~cm. It can also be observed that the error reaches its maximum around 1.00~m/s and then decreases slightly as the speed further increases. These results show that the proposed framework maintains stable localization accuracy over a range of operating speeds. This behavior can be attributed to the joint modeling of local temporal patterns, long range contextual dependencies, and sequential motion continuity. The results also indicate that the learned mapping from signals to positions can tolerate response variations caused by speed changes within the tested constant speed range.

Fig. \ref{fig_16} presents the localization performance under different vehicle load conditions, showing the influence of vehicle load variation. Compared with the unloaded condition (300~kg), only limited changes are observed in MAE, RMSE, and MaxAE under the loaded condition (500~kg). Although the load change alters the vibration response amplitude to some extent, the overall localization accuracy remains stable. This result suggests that the proposed framework is stable to moderate variations in AGV mass and possesses favorable adaptability for practical industrial deployment.

\begin{figure}[!t] 
\centering
\includegraphics[width=1\columnwidth, height=0.6\linewidth, keepaspectratio=false]{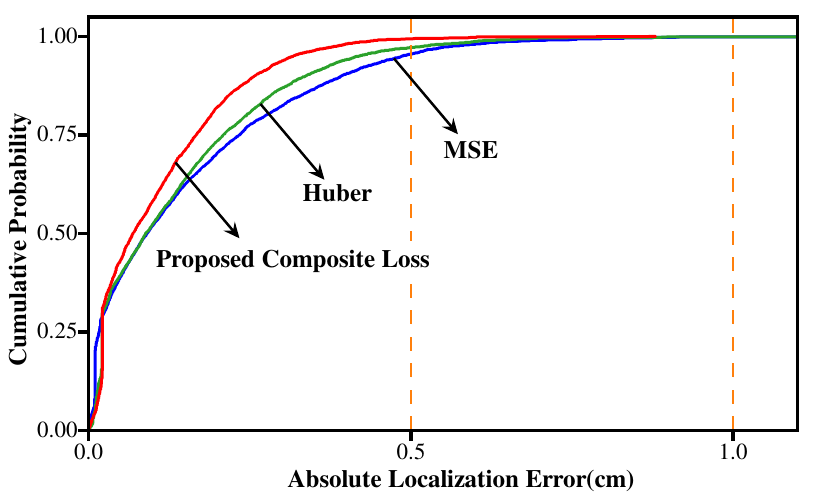}
\caption{Cumulative distribution comparison of absolute localization errors under different loss functions.}
\label{fig_17}
\end{figure}

\subsection{Ablation and Robustness Analysis}
In order to evaluate the effectiveness of the proposed deep learning framework, we conduct ablation studies from two aspects: network architecture and loss function design.

An ablation study on network architecture is conducted to evaluate the contributions of convolutional feature extraction, feature enhancement through self-attention, and temporal modeling through LSTM. For fair comparison, all models are trained and tested using the same dataset partition, preprocessing strategy, optimizer settings, training schedule, and the proposed composite loss. The compared architectures include a pure CNN model, a CNN-LSTM model, a CNN-Attention model, and the full CNN-Attention-LSTM model proposed in this work. The quantitative results are listed in Table \ref{tab:performance_comparison}.

\begin{table*}[!t]
\caption{Performance Comparison of Different Network Architectures}
\label{tab:performance_comparison}
\centering
\footnotesize

\begin{tabularx}{\textwidth}{YYYY}
\toprule
\textbf{Model} & \textbf{MAE (cm)} & \textbf{RMSE (cm)} & \textbf{MaxAE (cm)} \\
\midrule
CNN               & $0.761 \pm 0.073$ & $1.149 \pm 0.119$ & $6.286 \pm 0.765$ \\ [3pt]
CNN-LSTM          & $0.198 \pm 0.016$ & $0.280 \pm 0.020$ & $1.168 \pm 0.113$ \\ [3pt]
CNN-Attention     & $0.292 \pm 0.041$ & $0.418 \pm 0.058$ & $3.319 \pm 0.661$ \\ [3pt]
\midrule 
\textbf{Proposed Full Model} & $\mathbf{0.125 \pm 0.018}$ & $\mathbf{0.179 \pm 0.029}$ & $\mathbf{0.834 \pm 0.134}$ \\
\bottomrule
\end{tabularx}
\end{table*}

\begin{table*}[!t]
\caption{Performance Comparison of Different Loss Functions}
\label{tab:loss_performance_comparison}
\centering
\footnotesize

\begin{tabularx}{\textwidth}{YYYY}
\toprule
\textbf{Loss Function} & \textbf{MAE (cm)} & \textbf{RMSE (cm)} & \textbf{MaxAE (cm)} \\
\midrule
MSE                     & $0.140 \pm 0.005$ & $0.223 \pm 0.038$ & $1.928 \pm 1.247$ \\ [3pt]
Huber                   & $0.141 \pm 0.008$ & $0.215 \pm 0.014$ & $1.734 \pm 0.992$ \\ [3pt]
\midrule 
\textbf{Proposed Composite Loss} & $\mathbf{0.125 \pm 0.018}$ & $\mathbf{0.179 \pm 0.029}$ & $\mathbf{0.834 \pm 0.134}$ \\

\bottomrule
\end{tabularx}
\end{table*}

As shown in Table \ref{tab:performance_comparison}, the pure CNN model yields the weakest performance, indicating that local convolution alone is inadequate to fully characterize the spatiotemporal evolution of multichannel AGV vibration signals. After introducing the LSTM module, the localization error is significantly reduced, confirming the importance of temporal continuity modeling. The CNN-Attention model also outperforms the CNN baseline, showing that the self-attention mechanism can enhance feature representation by capturing long-range dependencies. Among all compared architectures, the full CNN-Attention-LSTM model achieves the best performance, demonstrating that self-attention for global dependency modeling and LSTM for sequential modeling are complementary for high precision AGV localization.

In addition, we implement an ablation study on loss function design to evaluate the effectiveness of the proposed composite loss. In this experiment, the network architecture is fixed as the full CNN-Attention-LSTM model, and only the loss function is changed for comparison. Table \ref{tab:loss_performance_comparison} summarizes the results obtained with three loss functions: the MSE loss, the standard Huber loss, and the proposed composite loss.

The results show that the MSE loss is more sensitive to large error samples and therefore is less effective in controlling extreme errors. Compared with MSE, the Huber loss reduces RMSE and MaxAE while maintaining a comparable MAE, indicating improved robustness to outliers during training. The proposed composite loss achieves the best overall performance, particularly in terms of MaxAE and large error suppression. That is because the additional excess error penalty explicitly suppresses samples with large errors while preserving accurate fitting for the majority of samples. Fig. \ref{fig_17} compares the error distributions under different loss functions. The proposed composite loss achieves a higher cumulative probability at the same error threshold than the MSE and Huber losses, indicating a more concentrated error distribution. Moreover, its curve approaches unity earlier, suggesting a shorter error tail and better suppression of large error samples. This observation is consistent with the quantitative results in Table \ref{tab:loss_performance_comparison}.

\section{Conclusion and Future Work}
\label{sec:Conclusion and Future Work}
This paper focuses on continuous high precision AGV localization in industrial environments with electromagnetic sensitivity and limited wireless communication. A shallowly buried interferometric FBG array sensing track is constructed to generate stable signals. Subsequently, a spatiotemporal feature representation is established through parametric modeling, task tailored preprocessing, and a sliding time window mechanism. On this basis, we design a hybrid deep learning network to capture the spatial correlation and temporal continuity of multichannel optical fiber signals.

Both simulation and experimental results confirm the effectiveness of the proposed localization method. The MAE values for the simulated and measured signals are 0.1278~cm and 1.9203~cm, respectively. In repeated experiments at the same speed as well as under varying speed and load conditions, the average localization error remains below 2~cm, demonstrating good accuracy, repeatability, and robustness.

Overall, the proposed method provides a feasible technical solution for continuous high precision AGV localization in wireless restricted industrial environments. Future work will further evaluate the proposed method under more complex driving profiles, including acceleration, deceleration, stop-and-go operation, and off track conditions. Further experiments will also examine the influence of different floor structures and sensing track installation conditions on signal transfer and localization performance, together with appropriate calibration or model fine-tuning strategies for new deployments \cite{ref43}, \cite{ref44}, \cite{ref45}. Joint longitudinal and lateral localization based on multi-line FBG sensing configurations will also be investigated \cite{ref46}.


\bibliographystyle{IEEEtran}
\bibliography{refs}

\vfill
\end{document}